\documentstyle[preprint,aps]{revtex}

\begin{document}

\preprint{EFUAZ\, FT-95-11}

\hsize = 7.0in
\widetext
\draft
\tighten

\title{Neutral Particles in Light of\linebreak
the Majorana-Ahluwalia Ideas\thanks{Submitted to
{\it ``International Journal of Theoretical Physics".} A part
of this work has been reported at the XVIII Oaxtepec Symposium
on Nuclear Physics. Oaxtepec, Mor., Mexico. January 4-7, 1995.}}

\author{Valeri V. Dvoeglazov\thanks{On leave of absence from
{\it Dept. Theor. \& Nucl. Phys., Saratov State University,
Astrakhanskaya ul., 83, Saratov\, RUSSIA.} Internet
address: dvoeglazov@main1.jinr.dubna.su}}

\address{Escuela de F\'{\i}sica, Universidad Aut\'onoma de Zacatecas \\
Antonio Doval\'{\i} Jaime\, s/n, Zacatecas 98000, ZAC., M\'exico\\
Internet address:  VALERI@CANTERA.REDUAZ.MX}

\date{February 28, 1995}

\maketitle

\begin{abstract}
\quad The first part of this article (Sections I and II)
presents oneself an overview of theory and phenomenology of
truly neutral particles based on the papers of Majorana, Racah, Furry,
McLennan and Case.  The recent development of  the construct, undertaken
by Ahluwalia [{\it Mod. Phys. Lett. A}{\bf 9} (1994) 439; {\it Acta Phys.
Polon. B}{\bf 25} (1994) 1267; Preprints LANL
LA-UR-94-1252, LA-UR-94-3118], could be relevant for explanation of
the present experimental situation  in neutrino physics and astrophysics.

In Section III  the  new fundamental
wave equations for  self/anti-self conjugate type-II spinors, proposed
by Ahluwalia, are re-casted to  covariant form. The connection
with the  Foldy-Nigam-Bargmann-Wightman-Wigner (FNBWW) type
quantum field theory is found. The possible applications to
the problem of neutrino oscillations are discussed.
\end{abstract}

\pacs{PACS numbers: 03.65.Pm, 11.30.Er, 14.60.Pq, 14.60.St}

%\maketitle

\renewcommand{\thefootnote}{\arabic{footnote}}

\newpage

\section{Introduction}

Neutrino physics and astrophysics brought many ``black spots" coming
from experiment at the  cloudless sky of the Standard Model.
{\it E.~g.}, Professor Robertson noted in this connection~\cite{Robert}:
``The solar neutrino results yield fairly strong and consistent
indications  that neutrino oscillations~\cite{Ponte,Gribov,osc} are
occurring.\footnote{There are opposite opinions on the solar neutrino
problems. {\it E.~g.},  in his  talk  ``The steady vanishing of the
three solar neutrino problems" at the 27th Int. Conf. on
High Energy Physics (1994) Prof. D. R. O. Morrison
denies~\cite{Mor}  their existence at all: ``The evidence
for any solar neutrino
problem is ``not compelling".}"  Though
``other evidence for new physics is  less consistent and convincing",
the solar neutrino  problem, ref.~\cite{Langa} (and in addition:
the ``negative mass squared" problem, {\it e.~g.},
ref.~\cite{Robert,mass}, the atmospheric neutrino anomaly~\cite{ANA},
the possibility of neutrinoless double $\beta$-
decay~\cite{beta,Rosen}, the ``spin crisis"
in QCD~\cite{EMC,Dor}, the tentative experimental evidence for a tensor
coupling in the $\pi^- \rightarrow e^- +\bar\nu_e + \gamma$
decay~\cite{decay}, as well as the dark matter problem,
{\it e.~g.},~\cite{dark})
seems to me to provide sufficient reasons for  searches of the models
beyond the framework of the Standard Model.  At the same time, the present
experimental situation does not provide clear hints for theoreticians, what
principles should be used for  explanation of the mentioned phenomena
and  for construction of the ``ultimate" theory.
Thus, the Nature leaves us with many ``degrees of freedom" of working out
the hypotheses which seem for the first sight to be very
``exotic"~\cite{Joshi,mirror,Bandy2,DVA0,tach}, if not
``crazy"~\cite{DVA-old,DVO}.

In this essay I  continue the study of $j=1/2$ and $j=1$ neutral
particles (the present-of-day knowledge states that neutrino
and photon are the only truly neutral particles in the Nature),
undertaken in ref.~\cite{DVA}. The crucial point of those papers
is based on  realizing   ``the dynamical role played by space-time
symmetries for [fundamental] interactions". The {\it ab initio}
construction of self/anti-self conjugate spinors in the $(j,0)\oplus
(0,j)$ representation space and derivation of some physical relevant
properties connected with space-time symmetries were presented there.
In fact, the articles~\cite{DVA} are the development of the formalism
proposed in the old papers~\cite{Majorana,Racah,Furry,Serpe,MCL,Case} and
they could be applicable for description
of neutrino interactions and clarification of
the present experimental situation.

\section{Theory and Phenomenology of Neutral Particles}

Kayser~\cite{kay-com} writes: ``We have become accustomed to thinking
of a neutrino $\nu$ and its antineutrino $\bar \nu$ as distinct
particles.\footnote{Thanks to the two-component  neutrino theory
proposed by Landau, Lee, Yang and Salam~\cite{LLYS}.}
However, it has long been recognized that the apparent
distinction between them may be only an illusion.  [Such] models,
[in which there is no difference between neutrino and
its antineutrino], naturally follow from GUT (grand unification theories)."
Moreover, from a viewpoint of a lot of models beyond the Standard one
it is very natural for neutrino (the spin-1/2 truly neutral particle) to be
a massive\footnote{Surprisingly, the six of the present upper bounds
on $m_{\nu_e}^2$
are negative~\cite{rev}.  {\it E.~g.},
the LANL result is $-147 \pm 68 \pm 41 \, eV^2$,
the LLNL one, $-130\pm 20 \pm 15\, eV^2$. The  most recent  measurement
(Troitsk, 1994) involves a new kind of systematics and
gives $-18\pm 6\, eV^2$.}
(as opposed to the Glashow-Salam-Weinberg electroweak model).

For the moment I take a liberty to present a little of history.
In 1937 Majorana
has  given a derivation of a symmetrical theory of
the electron and the positron~\cite{Majorana}.
The essential ingredient of that theory was the reformulation of
the variational principle,  based on the use of non-commutative variables.
This led him to  separation of the  Dirac equation
``into two distinct groups
one of which acts on the real part and the other, on the imaginary part
of [the spinor wave function], $\Psi = U +iV$\,\,". He noted: ``...
the part of this formalism which refers to the $U$ (or to the $V$) may be
considered by itself as a theoretical description of some material system,
in conformity with the general methods of quantum mechanics...
Equations constitute the simplest theoretical representation of
a system of neutral particles."
His ideas have been   developed
in  application to  the $\beta$ radioactivity
by Racah~\cite{Racah} and Furry~\cite{Furry}.
In fact, they have analyzed the Majorana's
projection\footnote{The notation of ref.~\cite{DVA}
is used through the present paper, which is different from
the papers~\cite{Racah,Furry,MCL,Case,Ryan}.}
\begin{equation}\label{proj}
\psi \rightarrow {1\over 2} \left \{ \psi +  S^c_{[1/2]} \psi \right
\}\quad .
\end{equation}
The matrix of  charge conjugation is defined as
\begin{equation}
S^c_{[1/2]}
\,=\,
e^{i\vartheta^c_{[1/2]}}
\,
\left(
\begin{array}{cc}
0 & i\,\Theta_{[1/2]}\\
-\,i\,\Theta_{[1/2]} &0
\end{array}\right)\,{\cal K}\,\equiv\,{\cal C}_{[1/2]}\,{\cal K}
\quad, \label{cph}
\end{equation}
where ${\cal K}$ is the operation of complex conjugation and
\begin{equation}\label{Wigner}
\left (\Theta_{[j]} \right )_{\sigma , \, \sigma^\prime}
= (-1)^{j+\sigma} \delta_{\sigma^\prime ,\,  -\sigma}
\end{equation}
is the Wigner's  operator ($\Theta_{[j]} {\bf J} \Theta^{-1}_{[j]}
= - {\bf J}^*$).
Racah noted that the symmetric description of a particle
and an antiparticle does not always imply that two types of particle are
physically undistinguishable.  That is clear for the electron
and the positron states, which have opposite electric charge,
but this statement can also be applied for neutrino:
``a neutrino emitted in a $\beta^-$ process may by absorption
induce only a $\beta^+$ process, and vice versa".
However, if consider the symmetric Hamiltonian (the sum of $H_F$,
the Fermi Hamiltonian,
and  $H_{KU}$, the Konopinski-Uhlenbeck Hamiltonian, ref.~\cite{FKU}),
we come  to
the physical identity between neutrino and antineutrino and, hence,
to the Majorana's
formalism for neutral particles, -- Racah writes,
from what follows the experimental
possibility of the  neutrinoless double $\beta$ decay discussed below.
In the papers~\cite{Furry} the Lorentz invariance  of
the Majorana's projection
(\ref{proj}) as well as the persistence in  time and the possibility of
interaction of the Majorana's particle with
the nonelectric scalar potential $\gamma_0 \Phi$ had been proven.
Furry also noted the non-invariance of the projection
under the change of phase  ({\it i.~e.}, in fact,
with respect to multiplication by a complex constant,
what implies the absence of the simple gauge interactions
of the Majorana neutral particle as opposed to the Dirac charged particle).
Differing from Racah he has claimed that  ``the results
predicted for ... observed processes [$\beta$-radioactivity]
are ... identical   with
those of the ordinary theory.  [However],  the physical
interpretation is quite different  [and]
an experimental decision between the formulation using
neutrinos and antineutrinos
and that using only neutrinos will ... be ... difficult.\footnote{Of course,
in the case of
massless states this assertion does not cause any  opposite opinions. Also,
in ref.~\cite{Ryan} the equivalence of description of the neutrino
in terms of Majorana spinors and Weil spinors was claimed,
but let us not forget that their arguments
implied zero neutrino mass.  I would like to mention the very
detailed  pedagogical introduction of ref.~\cite{Mannheim} to  the Majorana
theory, which has included a discussion of mass eigenstates of  the neutrino.
Nevertheless, in the case of  massive neutrinos
more explanations are required
to the problem of the equivalence of the two descriptions and
to the question of  the number of  independent states.
See also the footnote \# 22 in ref.~[25a], ref.~\cite{DVA,Sokolov}
and the discussion below."} His point
of view  is now widely accepted:   as opposed to the Dirac prescription  of
the charged particle (that has four states which answer for the same momentum
but different
spin configurations of particle and antiparticle)  in the Majorana
theory for $j=1/2$ particles there are just two states
corresponding to the two projections of the spin,  {\it i.~e.} there are no
``antiparticles" and  any necessity of the negative-energy states.

Important reformulations of  the Majorana's work  have
been undertaken by McLennan and Case in 1957, ref.~\cite{MCL,Case}.
Let me reproduce the main points of the Case's paper\footnote{The papers of
Serpe~\cite{Serpe} and McLennan~\cite{MCL} are concerned with
the massless neutrino and could be accounted as  the
particular cases. Let us not
forget that we don't have a strong theoretical principle that forbids
mass of neutrino.}.
By using Majorana {\it ansatz}\footnote{Let us note that the definitions
of K. M. Case and D. V. Ahluwalia differ by the overall phase factor.}
\begin{equation}
\psi_{L} = {\cal C}^{-1}_{[1/2]} \psi_{R}^* \quad ,
\end{equation}
where
$\psi_{R,L} ={1\over 2} (1\pm \gamma_5)\psi$,  the Dirac equation was shown
to be re-written to
\begin{equation}
\eta^\mu \partial_\mu \phi +\kappa \phi^* = 0\quad ,
\end{equation}
and its complex conjugated
\begin{equation}
\eta^{\mu\,*}\,\partial_\mu \phi^* +\kappa \phi =0 \quad.
\end{equation}
Here $\eta^\mu = {\cal C}_{[1/2]} \gamma_{-}^\mu={\cal C}_{[1/2]}
(1-\gamma_5)\gamma^{\mu}/2$,
\,\, $\phi = \psi_{R}$ and $\kappa$
is mass of the particle in the notation of ref.~\cite{Case}.
The matrices $\eta^\mu$ satisfy the anticommutation relation:
\begin{equation}
\eta^{\mu\,*} \,\,\eta^\nu +\eta^{\nu\,*}\,\,\eta^\mu = 2g^{\mu\nu}\quad .
\end{equation}
The signature was chosen to be $(-1, +1, +1, +1)$.
The corresponding Hamiltonian equations are
\begin{eqnarray}
i\frac{\partial \phi}{\partial t}
&=& {1\over i} \, \bbox{\sigma} \cdot \bbox{\nabla}\phi +
\kappa (A \phi^*)\quad, \\
i\frac{\partial (A\phi^*)}{\partial t}
&=& - {1\over i} \, \bbox{\sigma} \cdot \bbox{\nabla} (A\phi^*)
+ \kappa  \phi\quad ,
\end{eqnarray}
with $\eta^\mu = -iA\sigma^\mu$. The matrix  $A$ can be chosen
$\sigma_2$ in the conventional representation (see~\cite[p.308]{Case}).
The law of association for the proper Lorentz transformation is usual,
$\Lambda = \exp \left  ({1\over 2} \, v \, \bbox{\sigma} \cdot {\bf q}\right )$
with velocity $v$ in the direction ${\bf q}$. However, for
spatial reflections one has to impose
\begin{equation}
\phi^\prime (x^\prime) = \Lambda \phi^* (x),
\quad \mbox{or}\quad \phi^* (x) =\Lambda^{-1} \phi^\prime (x^\prime)\quad .
\end{equation}
This form ensures that $\Lambda = i\rho A$, where $\rho$ is a real number
with the absolute value unity.
By  using similar arguments for time reflections one has
$\phi^\prime (x^\prime)= \Lambda \phi^* (x)$ where
$\Lambda =\mu A$, with $\mu$ being real (and its absolute value
being equal to the unit). However, the McLennan-Case
consideration does not exhaust all possible Majorana-like construct.
For instance,
the possibility of the anti-self conjugate construct, {\it i.~e.},
\begin{equation}
\psi \rightarrow {1\over 2} \left \{ \psi -  S^c_{[1/2]} \psi \right \}
\end{equation}
has been realised much later~\cite{Mannheim}.
{}From a physical point of view this corresponds to the
two neutrino with  opposite  $CP$ quantum numbers, {\it e.~g.}~\cite{Wolf}.

Recently, the theory of neutral Majorana-like particles
has been developed substantially in
the papers of Ahluwalia~\cite{DVA}.
Particularly, the generalization to
higher-spin particles has been proposed. The formalism is
based on the type-II bispinors (another Majorana-like  construct
which could be  important for description of higher spin particles)
introduced by him.  The   fundamentally new wave equation has been proposed
there. We are going to discuss it  in the next Section.

The type-II  $(j,0)\oplus (0,j)$  bispinors are defined
in the following way:
\begin{equation}\label{sp-dva}
\lambda(p^\mu)\,\equiv
\left(
\begin{array}{c}
\left ( \zeta_\lambda\,\Theta_{[j]}\right )\,\phi^\ast_{_L}(p^\mu)\\
\phi_{_L}(p^\mu)\\
\end{array}
\right)\,\,,\quad
\rho(p^\mu)\,\equiv
\left(
\begin{array}{c}
\phi_{_R}(p^\mu)\\
\left ( \zeta_\rho\,\Theta_{[j]}\right )^* \,\phi^\ast_{_R}(p^\mu)
\end{array}
\right)
\,\,\quad .\label{os}
\end{equation}
$\zeta_\lambda$ and $\zeta_\rho$ are phase factors that
are  fixed by the conditions of  self/anti-self conjugacy:
\begin{equation} S^c_{[1/2]}
\,\lambda(p^\mu)\,=\,\pm\,\lambda(p^\mu)\,\,,\quad S^c_{[1/2]}
\,\rho(p^\mu)\,=\,\pm\,\rho(p^\mu)\quad,\label{sac}
\end{equation}
for a $j=1/2$ case; and
\begin{equation}
\left[\Gamma^5\,S^c_{[1]}\right]\,
\lambda(p^\mu)\,=\,\pm\,\lambda(p^\mu)\,\,,\quad
\left[\Gamma^5\,S^c_{[1]}\right]\,
\rho(p^\mu)\,=\,\pm\,\rho(p^\mu)\quad,\label{sacc}
\end{equation}
for a $j=1$ case\footnote{The self/anti-self conjugate type-II
spinors were shown in  ref.~\cite{DVA}  not to exist for bosons.
This fact is related with the FNBWW-type construct and it follows
from the analysis of ref.~[18a].
However, $[\Gamma^5 S^c]$ self/anti-self
conjugate type-II spinors have been introduced there.}.
The spin-1 counterpart of the equation (\ref{cph}) is
\begin{equation}
S^c_{[1]}\,=\,
e^{i\vartheta^{c}_{[1]}}
\,
\left(
\begin{array}{cc}
0 & \Theta_{[1]}\\
-\,\Theta_{[1]} &0
\end{array}\right)\,{\cal K}\,\equiv\,{\cal C}_{[1]}\,{\cal K}\quad.
\label{cpo}
\end{equation}
The phase factors are determined as
$\zeta^S_\lambda\,=\, \zeta^S_\rho \,=\,+\,i$, for the self charge
conjugate $j=1/2$ spinors, $\lambda^S(p^\mu)$ and
$\rho^S(p^\mu)$; and
$\zeta^A_\lambda\,=\, \zeta^A_\rho\, =\, -\,i$, for
the anti-self charge conjugate
$j=1/2$ spinors, $\lambda^A(p^\mu)$ and $\rho^A(p^\mu)$.
The equations (\ref{sacc})\,
determine $\zeta^S_\lambda\,=\, \zeta^S_\rho\, = \, +\,1$ for the self
$\left[\Gamma^5\,S^c_{[1]}\right]$-conjugate  $j=1$ spinors; and
$\zeta^A_\lambda\,=\, \zeta^A_\rho\, = \, -\,1$ for the anti-self
$\left[\Gamma^5\,S^c_{[1]}\right]$-conjugate $j=1$
spinors.  The remarkable property of
the self/anti-self conjugate  spinors,
which seems not to be realised before an appearance of the papers~\cite{DVA},
is: they cannot be in the definite helicity eigenstates. In fact,
let the 2-spinors $\phi^{h}_{_{L,R}}({p}^\mu)$
be an eigenstate of the helicity
operator
\begin{equation}
{\bf J}\cdot\widehat{\bf p}\,\,
\phi^{h}_{_{L,R}}({p}^\mu)\,=\,h\, \phi^{h}_{_{L,R}}({p}^\mu)\quad,
\label{ida}
\end{equation}
then, by using the Wigner-identity (see formula  above,
Eq. (\ref{Wigner})), we convince ourselves
\begin{equation}
{\bf J}\cdot\widehat{\bf p}\,\,
\Theta_{[j]}\,\left[\phi^{h}_{_{L,R}}({p}^\mu)\right ]^\ast\,=\,-\,h\,
\Theta_{[j]}\,\left[\phi^{h}_{_{L,R}}({p}^\mu)\right]^\ast
\quad.\label{idb}
\end{equation}
Thus, if $\phi^{h}_{_{L,R}}({p}^\mu)$ are eigenvectors of ${\bf J}
\cdot\widehat{\bf p}$, then
$\Theta_{[j]}\,\left[\phi^{h}_{_{L,R}}({p}^\mu)\right]^\ast$ are
eigenvectors of ${\bf J} \cdot\widehat{\bf p}$ with {\it opposite}
eigenvalues to those associated with
$\phi^{h}_{_{L,R}}({p}^\mu)$, ref.~[22c,d].
The unusual properties of the type-II spinors under  space (time) reflections
have also  been  noted in ref.~\cite{DVA}.
They are not eigenspinors of the parity operator,
see formulas (36a,b) and (37a,b) in the fourth paper.

The key test for a Majorana neutrino is the neutrinoless double-beta
decay. An antineutrino emitted in the beta decay of one neutron
is supposed to interact with another neutron and to cause it to transform
into a proton and a electron. So in the final state there are two
protons, two electrons and no neutrinos, $(A,Z) \rightarrow (A, Z+2) +2e^-$.
The conservation of lepton number
is violated.  Such a possibility, originally proposed
by Racah~\cite{Racah}, did not yet observe in experiment
in spite of the fact that
the available phase space for this process is larger
than for the two-neutrino double  $\beta$ decay~\cite{Furry}.
The experimental  bound for a halftime of  neutrinoless $\beta$ decay  is
$T_{1/2} > 2\times 10^{24}$ years (the enriched isotope $^{76} Ge$ was used,
ref.~\cite{beta}).
The failure of its observation
was explained by the statement that apart from the
non-conservation of lepton number the Racah processes is inhibited by
helicity. In order to complete the second step of the Racah process,
the antineutrino has to flip  its helicity and turns itself into a
neutrino\footnote{Of course, this explanation is appropriate
only in the framework of the Standard Model.}.
Rosen has shown~\cite{Rosen} that such a flip may be induced
only by a Majorana mass term. ``... Even if right-handed
currents provide the
phenomenological mechanism for  no-neutrino decay, the
fundamental mechanism underlying the process  must be  [a presence of]
neutrino mass [term]."
In the case of neutral particles  the electric charge conservation
(superselection rules) no longer forbids transitions
between particle and antiparticle $\nu_{eL}
\leftrightarrow \bar \nu_{eR}$ or $ \bar \nu_{eL} \leftrightarrow \nu_{eR}$.
It is these oscillations that provide the ground for the Racah process.
For the first time a theoretical model of neutrino oscillations
has been proposed by Pontecorvo in 1957, ref.~\cite{Ponte}\footnote{As
mentioned in ref.~\cite{Rosen}, some rumors of the positive result
concerning no-neutrino decay were circulated in the end of the fifties.},
see also~\cite{Maki}, by using the analogy with oscillations in
the $K^0 - \bar K^0$  spinless meson system~\cite{KK}.
This old idea had eventually been gone of the use. But it has found
a new life in the idea of oscillations between
different flavours~\cite{Gribov,osc,Ponte2,other}
in connection with the discovery of muon and $\tau$-lepton neutrinos.

Since in the third Section\,\, I am going to deal with a  scheme of
neutrino oscillations
on the ground of Majorana-like  theory with type-II spinors
let me reproduce here main points of  the well-known flavour
mixing scheme\footnote{More extended consideration could  be
found in~\cite{osc,Hughes,Rolnick}.} and  of
the common-used consideration of neutrino mass terms.

Schemes of neutrino mixing are usually
characterized by the type of the relevant mass term.
According to the modern literature it is possible to form
the following mass terms in the Lagrangian\footnote{The present
experimental data restrict the number of light neutrino species
to three (electron, muon and~$\tau$-lepton neutrino~\cite{spec}).}:
\begin{itemize}
\item
Dirac mass term:
\begin{equation}
{\cal L}^D = - \sum_{l^\prime,\,l = e,\mu,\tau\ldots}
\bar \nu_{l^\prime\,R} M_{l^\prime l} \nu_{l\,L} +
h. c. \quad ;
\end{equation}
\item
Majorana mass term (left-left):
\begin{equation}
{\cal L}^M =-{1\over 2} \sum_{l^\prime,\, l=e,\mu\,\tau\ldots}
\overline{(\nu_{l^\prime \,L})^c} M_{l^\prime l} \nu_{l\,L}
+ h. c.\quad ;
\end{equation}

\item
Dirac plus Majorana mass term
\begin{equation}
{\cal L}^{D+M} = -{1\over 2} \sum_{l^\prime,\, l=e,\mu,\tau\ldots}
\overline{(\nu_{l^\prime \, L})^c}
M_{l^\prime l}^L \nu_{l\,L} \, - \sum_{l^\prime ,\,l=e,\mu,\tau\ldots} \bar
\nu_{l^\prime\,R} M_{l^\prime l}^D \nu_{l\,L} \, - \, {1\over 2}
\sum_{l^\prime,\,l =e,\mu,\tau\ldots}
\overline{\nu_{l^\prime \,R}} M_{l^\prime l}^R
(\nu_{l\,R})^c + h. c.
\end{equation}
\end{itemize}
So, in a general case
it is necessary to consider three (six) mass eigenstates that correspond
to the  diagonalized mass matrix obtained by the unitary transformation
with the $3\otimes 3$ (or $6\otimes 6$ in the case of the D+M mass term)
matrix, {\it e.~g.}, $\nu_{lL} = \sum_{i=1}^3 U_{li} \nu_{iL}$.
We will  denote
the mass eigenstates $\vert \nu_i>$, $i=1,2,3$.  Thus, one can obtain the
diagonalized mass term in the Lagrangian:
\begin{eqnarray}
{\cal L}^D &=&
- \sum_{i=1}^3 m_i \bar \nu_i \nu_i \quad,\\ {\cal L}^{M (D+M)} &=&
-{1\over 2} \sum_{i=1}^{3 (6)} m_i \bar \chi_i \chi_i\quad .
\end{eqnarray}
The most general mass matrix  (Dirac and Majorana mass term)
can be represented  in the following  form:
\begin{eqnarray}
\lefteqn{\overline{\Psi_{L,R}} M \Psi_{L,R}= }\nonumber\\
&&\pmatrix{ \bar \psi_L  &  (\bar\psi_R)^c & (\bar\psi_L)^c
& \bar \psi_R\cr}
\pmatrix{ 0 & 0 & m_L & m_D\cr
0 & 0 & m_D & m_R\cr
m_L & m_D & 0 & 0\cr
m_D & m_R & 0 & 0\cr}
\pmatrix{\psi_L\cr (\psi_R)^c\cr (\psi_L)^c \cr \psi_R\cr}\quad.
\end{eqnarray}

In the vacuum mass eigenstates  propagate independently,
{\it i.~e.} let assume that they are orthogonal.
If  a physical state  is the linear combination of mass eigenstates which
have different masses  (for the sake of simplicity
let me consider only two species) one has:
\begin{eqnarray}\label{sup}
\cases{\vert \nu_e (0) > = cos \theta_\nu \,\vert \nu_1 >
+ sin\theta_\nu  \,\vert \nu_2 > &\cr
\vert \nu_\mu (0) > = -sin \theta_\nu \, \vert \nu_1 > + cos\theta_\nu
\, \vert \nu_2 >\quad; &}
\end{eqnarray}
the partial content of species in it   may  vary with time.
Let in the instant of time $t=0$  we have  the mixing (\ref{sup}), then
at a later time $t$
\begin{eqnarray}
\lefteqn{\vert \nu_e (t) > = cos\theta_\nu \, e^{-iE_1 t}\,\vert \nu_1 >
+\,sin\theta_\nu\,e^{-iE_2 t} \, \vert \nu_2 > =\nonumber}\\
&=& \left (e^{-iE_1 t} \, cos^2
\theta_\nu +e^{-E_2 t} \, sin^2 \theta_\nu\right ) \, \vert \nu_e (0)> +
\,sin\theta_\nu \, cos\theta_\nu \left (e^{-iE_2 t} - e^{-E_1 t}\right )\,
\vert
\nu_\mu (0)>\quad,
\end{eqnarray}
and
\begin{eqnarray}
\lefteqn{\vert \nu_\mu (t) > = -sin\theta_\nu \, e^{-iE_1 t}\, \vert
\nu_1 > +\,cos\theta_\nu \, e^{-E_2 t} \, \vert \nu_2 > =\nonumber}\\
&=& sin\theta_\nu \, cos\theta_\nu
\left (e^{-E_2t} - e^{-iE_1 t}\right ) \, \vert
\nu_e (0) >+  \left (e^{-iE_2 t} \, cos^2 \theta_\nu +e^{-E_1 t} \, sin^2
\theta_\nu \right )\, \vert \nu_\mu (0)>\quad .
\end{eqnarray}
Thus, a electron neutrino
produced at $t=0$ has non-zero probability of being a muon neutrino at a
later time (and vice versa).  The probability is calculated to give
\begin{eqnarray}\label{pem}
P_{\nu_e \rightarrow \nu_\mu} &=& \vert <\nu_\mu (0) \vert
\nu_e (t)> \vert^
2= \vert sin\theta_\nu \, cos\theta_\nu \left (e^{-iE_2 t} -
e^{-iE_1 t}\right )\vert^{\,2} =\nonumber\\
&=& 2\, sin^2 \theta_\nu \,cos^2
\theta_\nu \left [1-cos (E_1 -E_2) t \right ]\quad.
\end{eqnarray}
For the sake of completeness let us note that
\begin{equation}\label{pee}
P_{\nu_e \rightarrow \nu_e} = \vert <\nu_e (0) \vert
\nu_e (t)> \vert^2= 1-sin^{\, 2}\,  2\theta_\nu sin^2 \left [{1\over 2}
(E_2 -E_1)t\right ]\quad.
\end{equation}
Since in the high-velocity limit ($p >> m$)
\begin{equation}
E_1 -E_2 =\sqrt{p^2 +m_1^2}
-\sqrt{p^2 +m_2^2} \approx \frac{m_1^2 -m_2^2}{2p}\quad,
\end{equation}
one obtains
\begin{equation}
P_{\nu_e \rightarrow \nu_\mu} \approx 2\, sin^2\theta_\nu \,
cos^2\theta_\nu \left [\, 1 -cos \left (\frac{m_1^2
-m_2^2}{2p}\right ) \frac{c^3}{\hbar} t \, \right ]\quad,
\end{equation} where we
restored $c$ and $\hbar$ in order cosine to be dimensionless.  Since the
velocity of neutrino is approximately\, (?) \,  equal
to the light velocity, one has
\begin{eqnarray}
P_{\nu_e \rightarrow \nu_\mu} &\approx& 2\, sin^2\theta_\nu  \,
cos^2 \theta_\nu \left [ 1 -cos \left  ( \frac{m_1^2 -m_2^2}{2p} \right  )
\frac{c^2}{\hbar} x \right  ] =\nonumber\\
&=& 2\, sin^2 \theta_\nu \, cos^2
\theta_\nu \left [ 1 -cos \,\,2\pi \frac{x}{l_{12}}\right ]\quad,
\end{eqnarray}
where
\begin{equation}
l_{12} =\frac{4\pi p \hbar}{(m_1^2 -m_2^2) c^2}
\end{equation}
is the ``vacuum oscillation length". In the case of almost ``degenerate"
neutrinos
$(m_1^2 -m_2^2)\approx (10^{-2} eV/c^2)^2$ the ``oscillation  length"
$l_{12}$ is of the order of meters.
The readers are able to find  the numerous
literature on  the other versions of the oscillations
(including three species
etc.), see for the recent review~\cite{rev}.

The present-of-day experiments have not detected any such oscillations
for terrestrially (nuclear reactors, accelerators) created neutrinos.
That  is usually explained by very small mass differences
between eigenstates. On the other hand, the study of solar neutrinos
reveals a strong possibility that, before they reach
the Earth, the neutrinos undergo a significant oscillation. Besides
vacuum oscillations, plasma processes also should be taken into account
in the analysis of the solar neutrino flux.
However, we aren't going to discuss here the transmission through matter
(the Mikheyev-Smirnov-Wolfenstein effect~\cite{MSW}),
referring the reader to known reviews~\cite{reviews}.

For the moment, many physicists don't  consider  seriously
the Pontecorvo's original idea.  ``Since the helicity of a free particle
is conserved, in vacuum the oscillations $\nu_L \leftrightarrow \bar \nu_R$
cannot occur... For the above reasons it was generally
supposed that Pontecorvo's original oscillations are just
the oscillations of active neutrinos into sterile
states\footnote{{\it E.~g.}, $\nu_L \leftrightarrow \bar
\nu_L$.}, whereas the true neutrino-antineutrino
oscillations were considered impossible", -- claimed Akhmedov
{\it et al.}\footnote{{\it Cf.} the thoughts of refs.~\cite{Akhmed}
and~\cite[p.4,5]{Rosen} on neutrino-antineutrino oscillations.}
in ref.~\cite{Akhmed}.  Nevertheless, the same authors
realised that under certain conditions   particle-antiparticle
oscillations can occur and revisited the original idea
on the ground of introduction of
magnetic (or electric) dipole moment of neutrino with an addition of
neutrino of other specie.
The similar conclusion has been reached
in~\cite[p.378]{Hughes} where was said that ``traversal
of the solar magnetic field
may flip the neutrino spin." However, the estimated order
of  the transition magnetic moment  is $\mu_\nu \sim 10^{-11}
 - 10^{-10} \mu_B$.
``[Nevertheless], resonant effects in
a full treatment may well enhance the spin-flip to
a level where it is important."

\smallskip

\ldots It seems to me the history of the Majorana theory (as well as of
neutrino physics itself) is very dramatic:
one can see from the above that many outstanding physicists
were not able to find the common answers
on the experimental consequences
of this description\ldots

\smallskip

Next,  in the following Section we shall
work with spin-1 fields in the Weinberg formulation.
Therefore, it is useful   to repeat the key points of
this {\it particular}  model presented in the
papers~\cite{Weinberg,Sankar,Tucker,DVA0,DVO-old,DVO}.
The pioneer study of the $(j,0) \oplus (0,j)$  representation
space for description of higher spin particles
has been undertaken in ref.~\cite{Weinberg}.  This way of
a consideration is on an equal footing with the Dirac's way
of description of spin-1/2 particles and, in fact, has
its origin from the Wigner's classic work~\cite{Wigner}.
In the Weinberg theory a $2(2j+1)$ bispinor is constructed
from left- and right-spinors $\phi_R$ and $\phi_L$, with they
transforming according to the $(j,0)\oplus (0,j)$ representation
of the Lorentz group. Without reference to any wave equation it can be
shown that
\begin{eqnarray}
(j,\,0): &&\,\, \phi_{_R}( p^\mu)
\,=\, \Lambda_{_R}(p^\mu\leftarrow\overcirc{p}^\mu)
\,\phi_{_R}(\overcirc{p}^\mu)=
\exp\left (+{\bf J}\,\cdot\bbox{\varphi}\right)
\phi_{_R}(\overcirc{p}^\mu) \quad,\label{r}\\
(0,\,j): &&\,\,\phi_{_L}(p^\mu)\,
\,=\,\Lambda_{_L}(p^\mu\leftarrow\overcirc{p}^\mu)\,
\phi_{_L}(\overcirc{p}^\mu)=
\exp\left (-\,{\bf J}\,\cdot\bbox{\varphi}\right )
\phi_{_L}(\overcirc{p}^\mu)\quad,  \label{l}
\end{eqnarray}
where $\Lambda_L$ and $\Lambda_R$ are the Lorentz boost matrices for left-
and right  $j$- dimensional spinors from
the rest system $\overcirc{p}^\mu$; $\bbox{\varphi}$ are the Lorentz boost
parameters;
the operator ${\bf J}$
is presented by the angular momentum matrices.
The Weinberg equation contains solutions with tachyonic dispersion
relations\footnote{Let  us note that the  massless {\it first-order}
``Weinberg" equations for any spin have proven in
ref.~\cite[Table 2]{DVA-old} to possess another kinematical acausalities.
Apart from the correct physical dispersion $E=\pm p$ there is a wrong
dispersion relation $E=0$ in the case of $j=1$ (in the case of higher spins
one has even more  acausal solutions).
This fact doubts their application for all
processes (including quantumelectrodynamic
processes).  Nevertheless, the
massless limits of  the modified $2j$-order
Weinberg equations ($\wp_{u,v} = \pm 1$
for bosons) \begin{equation} \left [ \gamma^{\mu_1 \mu_2 \ldots
\mu_{2j}}\partial_{\mu_1}\partial_{\mu_2}\ldots\partial_{\mu_{2j}}
+\wp_{u,v} m^{2j}\right ]\Psi(x)=0,
\end{equation}
turn out to be well-defined and has no any kinematical
acausality~\cite{DVA-old}.
The $\gamma$- matrices are the covariantly defined $2(2j+1) \otimes
2(2j+1)$-matrices. See also refs.~\cite{DVO,DVO-old,DVO-old2}
for discussion on the connection of the Weinberg
formulation with the antisymmetric tensor field description and
for attempts of explanation of the origins
and the consequences of incorrect dispersion relations.}.
In  1971  Tucker and Hammer~\cite{Tucker}
have shown that it is possible
to reformulate the $2(2j+1)$ theory and to obtain
the spin-$j$ equations which possess
the {\it correct} physical dispersion.  Positive- and
negative-energy spinors  coincide in their construct.
However, introduction of electromagnetic gauge
interaction in their equation for $j = 1$ mesons
appears to be difficult.
The resulting theory is not renormalizable for all $j \geq 1$.
Another reformulation has been recently  (1993) proposed.
Based on the analysis of the transformation properties
left- and right- spinors and  a choice of appropriate rest spinors
(spinorial basis),
Ahluwalia {\it et al.}~\cite{DVA0}  have noted that it is possible
to construct the {\it Dirac}-like theory in  $(j,0)\oplus (0,j)$
space for  arbitrary spin $j$\footnote{See
also the old works of Sankaranarayanan and Good~\cite{Sankar}.}.
The remarkable feature of this construct is the fact that boson
and  its antiboson have opposite relative intrinsic parities.
Such  a type of  theory has been named as
the Foldy-Nigam-Bargmann-Wightman-Wigner (FNBWW) quantum field theory.
Finally, in my recent works~\cite{DVO}
another  Weinberg-Tucker-Hammer equation (``Weinberg  double")
with a correct physical dispersion has been given.
These equations turn out to be equivalent  to the equations for
the antisymmetric tensor $F_{\mu\nu}$ and  its dual,
which could be  deduced from the Proca theory.   The field  consideration
of the Weinberg doubles partly  clarified  contradictions with the
Weinberg theorem\footnote{The Weinberg theorem says that for
massless particles $B-A = helicity$,
if field transforms on the $(A,B)$  Lorentz group representation.},
occurred in the  earlier works~\cite{Hayashi,DVO-old}.
The contradictions were
caused by the application of the generalized Lorentz condition
(formulas (18) of ref.~[53a]) to physical quantum states what
resulted in equating the eigenvalues of the Pauli-Lyuban'sky
operator to zero.
The propagators for the Weinberg-Tucker-Hammer construct have also been
obtained, ref.~[21c].

However, these new constructs
deal with the Dirac-type spinors
(type-I spinors) and  they are applicable
mainly to the charge particles. Many questions related to
neutral particles have left unsolved in ref.~\cite{DVA0,DVO}.

\section{New Fundamental Equation Proposed by Ahluwalia and
Relevant Physical  Consequences}

The general wave equation for any spin in the instant-front formulation
of QFT is given in~[22c,d]\footnote{See the corresponding equation in
the light-front formulation in ref.~[22a].}
\begin{eqnarray}
\pmatrix{-\,\openone & \zeta_\lambda\,\exp\left(
{\bf J}\,\cdot \bbox{\varphi}\right )
\,\Theta_{[j]}\,{\mit\Xi}_{[j]}\, \exp\left( {\bf J}\,\cdot \bbox{\varphi}
\right )\cr
\zeta_\lambda\,\exp\left(-\, {\bf J}\,\cdot\bbox{\varphi}\right)
\,{\mit\Xi}^{-1}_{[j]}\,\Theta_{[j]}\,
\exp\left(- \,{\bf J}\,\cdot\bbox{\varphi}
\right) & -\,\openone}\,\lambda(p^\mu)\,=\,0.\label{genweq}
\end{eqnarray}
The particular cases ($j=1/2$ and $j=1$) are also given
there (Eqs. (31) and (32),
respectively). The $\lambda^S(p^\mu)$ appear to be the positive
energy solutions with $E\,=\,+\,\sqrt{m^2\,+\,{\bf{p}}^2}\,$,
the $\lambda^A(p^\mu)$, negative energy solutions with
$E\,=\,-\,\sqrt{m^2\,+\,{\bf{p}}^2}\,$ for both spin-1/2 and spin-1 cases.
However, to re-write these equations to  a covariant form
is a difficult task. For instance, an attempt of
the author of the formalism~\cite{DVA}
to put  the equation in the form $(\lambda^{\mu\nu} p_\mu p_\nu +
m \lambda^\mu p_\mu
-2m^2 \openone) \lambda (p^\mu)= 0$ was in a certain sense
misleading. He noted himself: ``it turns out that
[matrices] $\lambda^{\mu\nu}$ and $\lambda^\mu$ do not transform as
Poincar\'e tensors." Below I try  to explain in what
way the equations for $\lambda (p^\mu)$ and $\rho (p^\mu)$ spinors
are re-written to a covariant form.

The crucial point of derivation of the equation (\ref{genweq})
is the generalized Ryder-Burgard relation for type-II
spinors\footnote{In ref.~\cite{DVA0}
the relation $\phi_R (\overcirc{p}^\mu) = \pm \phi_L (\overcirc{p}^\mu)$
for type-I spinors
(in fact, for the Dirac bispinor) has been named
as the Ryder-Burgard relation,
see also~\cite[p.44]{Ryder}.  Through this paper I also use this name,
but I understand that this relation could be found in earlier
papers and books, see, {\it e.~g.}, the discussion
surrounding equations (25,26) of Ch.5, ref.~\cite{Novozh}. It can be
deduced also from  Eq. (22a) of ref.~\cite{Faustov}.}:
\begin{equation}
\left[\phi^{h}_{_{L}}({\overcirc{p}}^\mu)\right]^\ast\,=\,
{\mit\Xi}_{[j]}\,\phi^{h}_{_{L}}({\overcirc{p}}^\mu)\quad , \label{br}
\end{equation}
where
\begin{equation}
{\mit \Xi}_{[1/2]}\,=\,\left(\begin{array}{cc}
e^{i\phi} & 0\\
0 & e^{-i\phi}
\end{array}\right)\quad,\qquad
{\mit\Xi}_{[1]}\,=\,\left(\begin{array}{ccc}
e^{i\,2\,\phi}&0 & 0\\
0&1&0\\
0 & 0 & e^{-i\,2\,\phi}
\end{array}\right)\quad,
\end{equation}
$h$ is the helicity, $\phi$ is the azimutal angle associated
with ${\bf p}$.
In this framework ($j=1/2$ case) the best, what can be done,
is to re-write Eq. (\ref{genweq}) to the form:
\begin{equation}\label{rr}
\left (\frac{i\zeta_\lambda}{sin\theta}\gamma_5 \left [\bbox{\gamma}
\times \widehat{{\bf p}}\right ]_3
+\openone \right )\lambda (p^\mu)=0,
\end{equation}
($\theta$ is the polar angle associated with ${\bf p}$)
by using the following identities:
\begin{equation}
\Theta_{[1/2]}\Xi_{[1/2]} = \Xi^{-1}_{[1/2]}
\Theta_{[1/2]} = i\sigma_1 sin\phi - i\sigma_2 cos\phi = i\frac{\left
[\bbox{\sigma} \times {\bf p}\right ]_3}{\sqrt{p_1^2 +p_2^2}}\quad,
\end{equation}
\begin{equation}
\left [\bbox{\sigma} \times {\bf p}\right ] (\bbox{\sigma}\cdot {\bf p})
= - (\bbox{\sigma} \cdot{\bf p}) \left [\bbox{\sigma} \times {\bf p}\right ]
= i\bbox{\sigma} {\bf p}^2 -i{\bf p} (\bbox{\sigma}\cdot {\bf p})\quad,
\end{equation}
and
\begin{equation}
\exp (\pm \bbox{\sigma}\cdot\bbox{\varphi}/2) = cosh \,{\varphi \over 2}
\pm (\bbox{\sigma} \widehat{\bbox{\varphi}}) sinh \,{\varphi\over 2}, \quad
\widehat{\bbox{\varphi}} = \widehat{{\bf p}} = {{\bf p} \over \vert {\bf
p} \vert}\quad.
\end{equation}
However, the obtained equation can't be
considered as a dynamical equation (energy operator does not present
there). In fact, Eq. (\ref{rr}) is only a reformulation of the condition
of self/anti-self conjugacy.

Let us undertake another attempt.
{}From the analysis of the rest spinors (formulas 22a-22b and
23a-23c of ref.~[22d]) one can conclude that another form of
the generalized Ryder-Burgard relation is possible. Namely, the form
connecting 2-spinors of the opposite helicity is:
\begin{equation}\label{rbug12}
\left [\phi_L^h (\overcirc{p}^\mu)\right ]^* = (-1)^{1/2-h}
e^{-i(\theta_1 +\theta_2)}
\Theta_{[1/2]} \phi_L^{-h} (\overcirc{p}^\mu)\quad ,
\end{equation}
for a $j=1/2$ case;
and
\begin{equation}\label{rbug1}
\left [\phi_L^h
(\overcirc{p}^\mu)\right ]^* = (-1)^{1-h} e^{-i\delta}
\Theta_{[1]} \phi_L^{-h} (\overcirc{p}^\mu)\quad ,
\end{equation}
for a $j=1$ case ($\delta=\delta_1 +\delta_3$ for $h=\pm 1$ and
$\delta=2\delta_2$, for $h=0$).
Provided that the overall phase
factors of the rest spinors are chosen to be $\theta_1 +\theta_2=0$
(or $2\pi$) in a spin-1/2 case and
$\delta_1 +\delta_3 = 0 = \delta_2$, in a spin-1 case,
the Ryder-Burgard relation is written
\begin{equation}\label{rbu}
\left [\phi_L^h
(\overcirc{p}^\mu)\right ]^* = (-1)^{j-h} \Theta_{[j]} \phi_L^{-h}
(\overcirc{p}^\mu)\quad .
\end{equation}
This choice is convenient for calculations.
The same relations exist for right-handed spinors
$\phi_R (\overcirc{p}^\mu)$
in both a $j=1/2$ case and a $j=1$ case.

By using (\ref{rbu}) and following to the procedure of deriving
the wave equation developed in ref.~\cite{DVA} one can obtain
for a $j=1/2$ case ($\hat p=\gamma^\mu p_\mu$):
\begin{eqnarray}\label{eqq}
\left [{i\over m}\gamma_5\hat p - 1 \right ]
\Psi^{(S)}_{+1/2} (p^\mu) &=& 0\quad,\quad
\left [{i\over m}\gamma_5\hat p + 1 \right ] \Psi^{(A)}_{+1/2} (p^\mu) =0
\quad,\\ \label{eqq1}
&&\nonumber\\
\left [{i\over m}\gamma_5\hat p + 1 \right ]
\Psi^{(S)}_{-1/2} (p^\mu) &=& 0\quad,\quad
\left [{i\over m}\gamma_5\hat p - 1 \right ]
\Psi^{(A)}_{-1/2} (p^\mu) =0\quad.
\end{eqnarray}
Here we defined new spinor functions:
\begin{eqnarray}\label{sf}
\Psi^{(S)}_{+1/2} (p^\mu) = \pmatrix{i\Theta_{1/2}
\left [\phi_L^{-1/2} (p^\mu)\right ]^*\cr
\phi_L^{+1/2} (p^\mu)\cr}
\quad &\mbox{or}& \quad\Psi^{(S)}_{+1/2} (p^\mu)=
-i\pmatrix{\phi_R^{+1/2} (p^\mu)\cr
-i\Theta_{1/2} \left [\phi_R^{-1/2} (p^\mu)\right ]^*\cr},\\
\Psi^{(S)}_{-1/2} (p^\mu) = \pmatrix{i\Theta_{1/2}
\left [\phi_L^{+1/2} (p^\mu)\right ]^*\cr
\phi_L^{-1/2} (p^\mu)\cr}
\quad &\mbox{or}& \quad\Psi^{(S)}_{-1/2} (p^\mu)
= i\pmatrix{\phi_R^{-1/2}(p^\mu)\cr
-i\Theta_{1/2} \left [\phi_R^{+1/2} (p^\mu)\right ]^*\cr},\\
\Psi^{(A)}_{+1/2} (p^\mu) = \pmatrix{-i\Theta_{1/2}
\left [\phi_L^{-1/2} (p^\mu)\right ]^*\cr
\phi_L^{+1/2} (p^\mu)\cr}
\quad &\mbox{or}& \quad\Psi^{(A)}_{+1/2} (p^\mu)
= i\pmatrix{\phi_R^{+1/2} (p^\mu)\cr
i\Theta_{1/2} \left [\phi_R^{-1/2} (p^\mu)\right ]^*\cr},\\
\label{sfl}
\Psi^{(A)}_{-1/2} (p^\mu) = \pmatrix{-i\Theta_{1/2}
\left [\phi_L^{+1/2} (p^\mu)\right ]^*\cr
\phi_L^{-1/2} (p^\mu)\cr}
\quad &\mbox{or}& \quad\Psi^{(A)}_{-1/2} (p^\mu)=
-i\pmatrix{\phi_R^{-1/2} (p^\mu)\cr
i\Theta_{1/2} \left [\phi_R^{+1/2} (p^\mu)\right ]^*\cr}.
\end{eqnarray}
As opposed to $\lambda (p^\mu)$ and $\rho (p^\mu)$
these spinor functions
are the eigenfunctions of the helicity operator of
the $(1/2,0)\oplus (0,1/2)$ representation space, but they are not
self/anti-self conjugate spinors.

The equations similar to (\ref{eqq},\ref{eqq1}) can also
be obtained by the procedure
described in footnote \# 1 of ref.~[22d] with type-I
spinors ($\Psi=column (\phi_R (p^\mu)\quad \phi_L (p^\mu))$)
if imply that the Ryder-Burgard relation has the form
\begin{equation}
\phi_R (\overcirc{p}^\mu)=\pm i\phi_L (\overcirc{p}^\mu)\quad.
\end{equation}
The equations of kind (\ref{eqq},\ref{eqq1})
have been discussed in the old literature, ref.~\cite{Sokolik}.
Their relevance
to the problem of describing the neutrino has been noted in
the cited paper. The properties of this bispinors respective to
the parity ($\gamma_0$) operation are the following ({\it cf}. with
formulas (36a,b) in ref.~[22d]):
\begin{eqnarray}
\gamma_0 \Psi^{(S)}_{+1/2} (p^{\prime\,\mu}) &=& - i \left \{
\Psi^{(A)}_{-1/2} (p^{\mu})\right \}^c\quad,\\
\gamma_0 \Psi^{(S)}_{-1/2} (p^{\prime\,\mu}) &=& + i \left \{
\Psi^{(A)}_{+1/2} (p^{\mu})\right \}^c\quad,\\
\gamma_0 \Psi^{(A)}_{+1/2} (p^{\prime\,\mu}) &=& - i \left \{
\Psi^{(S)}_{-1/2} (p^{\mu})\right \}^c\quad,\\
\gamma_0 \Psi^{(A)}_{-1/2} (p^{\prime\,\mu}) &=& + i \left \{
\Psi^{(S)}_{+1/2} (p^{\mu})\right \}^c\quad.  \end{eqnarray}

By using the formulas relating $\Psi$, Eq. (\ref{sf}-\ref{sfl}),
with self/anti-self conjugate
spinors it is easy to find corresponding equations for spinors
$\lambda (p^\mu)$ and $\rho (p^\mu)$.
In the case of spin-1/2 field we obtain
\begin{eqnarray}\label{eql}
\hat p \lambda^S_{\uparrow} (p^\mu)+ im \lambda^S_{\downarrow} (p^\mu)
&=& 0\quad,\quad
\qquad \hat p \rho^S_{\uparrow} (p^\mu)- im \rho^S_{\downarrow} (p^\mu)
= 0\quad,\\
\hat p \lambda^S_{\downarrow}(p^\mu) - im \lambda^S_{\uparrow} (p^\mu)
&=& 0\quad,\quad
\qquad \hat p \rho^S_{\downarrow} (p^\mu)+ im \rho^S_{\uparrow} (p^\mu)
= 0\quad,\\
\hat p \lambda^A_{\uparrow} (p^\mu)- im \lambda^A_{\downarrow} (p^\mu)
&=& 0\quad,\quad
\qquad\hat p \rho^A_{\uparrow} (p^\mu)+ im \rho^A_{\downarrow} (p^\mu)
= 0\quad,\\ \label{eqll}
\hat p \lambda^A_{\downarrow} (p^\mu)+ im \lambda^A_{\uparrow} (p^\mu)
&=& 0 \quad,\quad
\qquad  \hat p \rho^A_{\downarrow} (p^\mu)- im \rho^A_{\uparrow} (p^\mu)= 0
\quad
\end{eqnarray}
(provided that $m\neq 0$). The indices $\uparrow$ or $\downarrow$
should be referred to the chiral helicity introduced
in~[22c,p.10].  If imply similarly to~[22d] that
$\lambda^S_{\uparrow\downarrow} (p^\mu)$ (and $\rho^A_{\uparrow\downarrow}
(p^\mu)$) are the positive-energy solutions and
$\lambda^A_{\uparrow\downarrow} (p^\mu)$
(and $\rho^S_{\uparrow\downarrow} (p^\mu)$)
are the negative-energy solutions, the equations (\ref{eql}-\ref{eqll})
in the coordinate space can be written
\begin{eqnarray}\label{cr1}
\partial_\mu \gamma^\mu \lambda_\eta (x) +
\wp_{\uparrow\downarrow} m\lambda_{-\eta} (x)&=&0\quad ,\\
\label{cr2}
\partial_\mu \gamma^\mu \rho_\eta (x) +
\wp_{\uparrow\downarrow} m\rho_{-\eta} (x)&=&0\quad ,
\end{eqnarray}
where $\wp_{\uparrow\downarrow}=\pm 1$ with the sign is ``$+$" if
$\eta=\uparrow$ and the sign is ``$-$" provided that $\eta=\downarrow$.
These equations (\ref{cr1}) and (\ref{cr2})
are very similar to the Dirac equation, however, the sign at the
mass term can be opposite and
spinors enter in the equations with
opposite chiral helicities. The Dirac equation with
the opposite sign at mass term had been considered (in different aspects)
in refs.~\cite{Markov,Beli,Brana}. Eqs. (\ref{cr1},\ref{cr2})
should be compared with the new form of
the Weinberg equation for $j=1$ spinors in a coordinate
representation, ref.~\cite{DVA0}.

One can incorporate the same chiral helicity states in equations
by using the identities (48a,b) of ref.~[22d].
\begin{eqnarray}\label{i1}
\rho^S_{\uparrow} (p^\mu) &=& - i\lambda^A_{\downarrow}(p^\mu)\quad,\quad
\rho^S_{\downarrow} (p^\mu) = + i\lambda^A_{\uparrow}(p^\mu)\quad,\\
\label{i2}
\rho^A_{\uparrow} (p^\mu) &=& + i\lambda^S_{\downarrow}(p^\mu)\quad,\quad
\rho^A_{\downarrow} (p^\mu) = - i\lambda^S_{\uparrow}(p^\mu)\quad.
\end{eqnarray}
Thus, one can come to
\begin{eqnarray}\label{sc1}
\hat p \lambda^S_{\uparrow\downarrow} (p^\mu)
+m \rho^A_{\uparrow\downarrow} (p^\mu)&=&0\quad,\quad
\hat p \lambda^A_{\uparrow\downarrow}(p^\mu)
+m \rho^S_{\uparrow\downarrow} (p^\mu) = 0\quad,\\ \label{sc2}
\hat p \rho^S_{\uparrow\downarrow}(p^\mu) +m
\lambda^A_{\uparrow\downarrow} (p^\mu) &=& 0\quad,\quad
\hat p \rho^A_{\uparrow\downarrow} (p^\mu)
+m \lambda^S_{\uparrow\downarrow}  (p^\mu) = 0 \quad .
\end{eqnarray}

It is also useful to note the connection of type-II spinors
$\lambda (p^\mu)$ and $\rho (p^\mu)$
with the type-I Dirac bispinor $\psi^D (p^\mu)$
and its charge conjugate $(\psi^D (p^\mu))^c$:
\begin{eqnarray}\label{con1}
\lambda^S (p^\mu) &=& \frac{1-\gamma_5}{2}
\psi^D (p^\mu)+\frac{1+\gamma_5}{2}(\psi^D (p^\mu))^c\quad,\\
\lambda^A (p^\mu)&=& \frac{1-\gamma_5}{2}
\psi^D (p^\mu)- \frac{1+\gamma_5}{2}(\psi^D(p^\mu))^c\quad,\\
\rho^S (p^\mu)&=& \frac{1+\gamma_5}{2}
\psi^D (p^\mu)+\frac{1-\gamma_5}{2}(\psi^D(p^\mu))^c\quad,\\ \label{conl}
\rho^A (p^\mu)&=& \frac{1+\gamma_5}{2}
\psi^D (p^\mu)- \frac{1-\gamma_5}{2}(\psi^D(p^\mu))^c\quad.
\end{eqnarray}
The equations (\ref{sc1},\ref{sc2})
could then be re-written to the form with type-I spinors:
\begin{eqnarray}
\left (\hat p +m \right ) \psi^D_{\pm 1/2} (p^\mu) +
\left (\hat p +m\right ) \gamma_5 (\psi^D_{\pm 1/2} (p^\mu))^c &=& 0\quad,\\
\left (\hat p -m \right ) \gamma_5\psi^D_{\pm 1/2} (p^\mu) -
\left (\hat p -m\right ) (\psi^D_{\pm 1/2} (p^\mu))^c &=& 0\quad,\\
\left (\hat p +m \right ) \psi^D_{\pm 1/2} (p^\mu) -
\left (\hat p +m\right ) \gamma_5 (\psi^D_{\pm 1/2} (p^\mu))^c &=& 0\quad,\\
\left (\hat p -m \right ) \gamma_5\psi^D_{\pm 1/2} (p^\mu) +
\left (\hat p -m\right ) (\psi^D_{\pm 1/2} (p^\mu))^c &=& 0\quad.
\end{eqnarray}
So, we can consider the $(\psi^D_h (p^\mu))^c$
(or $\gamma_5\psi^D_h (p^\mu)$, or their sum)
as the positive-energy solutions
of the Dirac equation and $\psi^D_h (p^\mu)$
(or $\gamma_5 (\psi^D_h (p^\mu))^c$, or their sum)
as the negative-energy solutions. The field operator can be defined
\begin{equation}\label{fo}
\Psi = \int \frac{d^3 {\bf p}}{(2\pi)^3} \frac{1}{2p_0}
\sum_{h} \left [(\psi^D_h (p^\mu))^c a_h exp (-ip\cdot x)
+ \psi^D_h (p^\mu) b_h^\dagger exp (ip\cdot x)\right ]\quad .
\end{equation}
The similar formulation has been developed by Nigam and
Foldy, ref.~\cite{Nigam}.

Let us note a interesting feature.
We can obtain the another interpretation (namely,
$\psi^D (p^\mu)$ corresponds
to the positive-energy solutions and $(\psi^D (p^\mu))^c$,
to the negative ones) if choose other overall phase factors
in the definitions of the rest-spinors $\phi_L (\overcirc{p}^\mu)$
and $\phi_R (\overcirc{p}^\mu)$,
formulas (22) of ref.~[22d].
The signs at the mass term depend on the form of the generalized
Ryder-Burgard relation; if $\theta_1 +\theta_2 =\pi$ the signs
would be opposite.  One can obtain the generalized equations
(\ref{eql}-\ref{eqll}) for an arbitrary choice  of the phase
factor. For $\lambda^S (p^\mu)$ spinors they are following:
\begin{eqnarray}
\cases{i\hat p \lambda^S_{\uparrow} (p^\mu)
- m {\cal T} \lambda^S_\downarrow  (p^\mu) =0&\cr
i\hat p \lambda^S_{\downarrow} (p^\mu) + m {\cal T}
\lambda^S_\uparrow (p^\mu) = 0,}
\end{eqnarray}
where
\begin{equation}
{\cal T} =\pmatrix{e^{i(\theta_1 + \theta_2)} &0\cr
0& e^{-i(\theta_1+\theta_2)}\cr}\quad
\end{equation}
and $m\neq 0$. In the case $\theta_1 +\theta_2 =\pm {\pi \over 2}$
we also have the correct physical dispersion,
$p_0^2 - {\bf p}^2 = m^2$, for $\lambda (p^\mu)$ spinors.

Next, one can see from (\ref{cr1},\ref{cr2})
that neither $\lambda^{S,A} (x)$ nor $\rho^{S,A} (x)$
are the eigenfunctions of the Hamiltonian operator (we have different
chiral helicities in the ``Dirac" equations).
They are not in mass eigenstates.
However $\psi^D$ and $(\psi^D)^c$ are in  mass and helicity eigenstates.
In ref.~\cite{Nigam} it was shown that even without a resort to a
plane-wave expansion, if  the eigenvector $\vert\phi >$ has the eigenvalue
``$-1$" of the normalized Hamiltonian $\hat H/\vert E\vert$ in the Hilbert
space, then $\vert\phi^c >$ has the eigenvalue ``$+1$".
This analysis is in accordance with the Feynman-St\"uckelberg
interpretation of ``antiparticle" as the particle moving backward in
time~\cite{FS}, which seems to be deeper with respect to the Dirac's hole
concept, because the former permits us to describe bosons on the equal
footing with fermions~\cite{DVA0}.
Thus, one can come to the conclusion that matrix
elements, {\it e.~g.}, $<\lambda^A_{-\eta} (0),\vert
\lambda^S_\eta (t)>$ have  the
non-zero value at the time $t$ ({\it cf.} with Eqs. (\ref{pem},\ref{pee})):
\begin{eqnarray}
< \lambda^A_{-\eta}\vert \lambda^S_\eta (t)> &\sim&  sin^2 ({Et\over
\hbar})\quad , \quad
< \lambda^S_{-\eta}\vert \lambda^S_\eta (t)> \sim  cos^2 ({Et\over
\hbar})\quad ,\\
< \lambda^S_{-\eta}\vert \lambda^A_\eta (t)> &\sim& sin^2 ({Et\over
\hbar})\quad , \quad
< \lambda^A_{-\eta}\vert \lambda^A_\eta (t)> \sim cos^2 ({Et\over
\hbar})\quad.
\end{eqnarray}
We are ready to put the question forward: can
the high-energy neutrino described by the field (Eq. (47)
of ref.~[22d])
\begin{equation}
\nu^{ML} \equiv \int \frac{d^3 {\bf p}}{(2\pi)^3} \frac{1}{2p_0}
\sum_{\eta}  \left [\lambda^S_\eta (p^\mu) a_\eta (p^\mu)  exp (-ip\cdot
x) + \lambda^A_\eta (p^\mu) a_\eta^\dagger (p^\mu) exp (ip\cdot x) \right ]
\end{equation}
``oscillate" from the state of one chiral helicity to another
chiral helicity with the oscillation length of the order
of the de Broglie wavelength,  $\lambda =h/p$ ?

For the case spin-1 the situation differs in some aspects.
Direct  calculations yield a non-dynamical quadratic (in
projections of the linear momentum)
equation:
\begin{equation}
\left [\zeta_{\lambda}\frac{\gamma_{11} p_2^2 +\gamma_{22} p_1^2
- 2\gamma_{12} p_1 p_2}{{\bf p}^2 -p_3^2} + \openone\right ]
\lambda (p^\mu) =0\quad.
\end{equation}
It can also be written in the form\footnote{We use the notation
in terms of the Barut-Muzinich-Williams matrices here,
ref.~\cite{Barut}.}:
\begin{eqnarray}
\pmatrix{- \openone & \zeta_{\lambda} D^{(1,0)} (i\frac{\left [
\bbox{\sigma}\times {\bf p}\right ]_3}{\sqrt{{\bf p}^2 - p_3^2}})\cr
\zeta_{\lambda} \Theta_{[1]} D^{(0,1)} (i\frac{\left [ \bbox{\sigma}\times
{\bf p}\right ]_3}{\sqrt{{\bf p}^2 - p_3^2}})\Theta_{[1]} & - \openone}
\lambda (p^\mu) = 0 \quad,
\end{eqnarray}
that is obtained by using, {\it e.~g.}, the technique
of ref.~\cite{Novozh}; $D^{(J,0)} (A)$ are the Wigner functions
for the $(J,0)$ representation, $D^{(0,J)} (A)$,
for the $(0,J)$  representation.

If accept another formulation of the Burgard-Ryder
relation (\ref{rbu}) one has\footnote{Again, one can obtain the
opposite signs in the equations if imply
$\delta_1 +\delta_3 =\pi$ \,\, for  $\phi_L (\overcirc{p}^\mu)$
and,  correspondingly, for $\phi_R (\overcirc{p}^\mu)$.}
\begin{eqnarray}
\gamma_{\mu\nu} p^\mu p^\nu \lambda^S_{\uparrow} (p^\mu) - m^2
\lambda^S_{\downarrow}  (p^\mu)&=& 0 \quad,\quad
\gamma_{\mu\nu} p^\mu p^\nu \rho^S_{\uparrow}  (p^\mu) - m^2
\rho^S_{\downarrow}  (p^\mu)= 0 \quad,\\
\gamma_{\mu\nu} p^\mu p^\nu \lambda^S_{\downarrow} (p^\mu)- m^2
\lambda^S_{\uparrow}  (p^\mu)&=& 0 \quad,\quad
\gamma_{\mu\nu} p^\mu p^\nu \rho^S_{\downarrow} (p^\mu)- m^2
\rho^S_{\uparrow}  (p^\mu) = 0 \quad,\\
\gamma_{\mu\nu} p^\mu p^\nu \lambda^S_{\rightarrow} (p^\mu) + m^2
\lambda^S_{\rightarrow}  (p^\mu)&=& 0 \quad, \quad
\gamma_{\mu\nu} p^\mu p^\nu \rho^S_{\rightarrow}  (p^\mu) + m^2
\rho^S_{\rightarrow}  (p^\mu)= 0 \quad,\\
\gamma_{\mu\nu} p^\mu p^\nu \lambda^A_{\uparrow} (p^\mu) + m^2
\lambda^A_{\downarrow} (p^\mu) &=& 0 \quad,\quad
\gamma_{\mu\nu} p^\mu p^\nu \rho^A_{\uparrow} (p^\mu) + m^2
\rho^A_{\downarrow} (p^\mu)= 0 \quad,\\
\gamma_{\mu\nu} p^\mu p^\nu \lambda^A_{\downarrow} (p^\mu)+ m^2
\lambda^A_{\uparrow} (p^\mu)&=& 0 \quad,\quad
\gamma_{\mu\nu} p^\mu p^\nu \rho^A_{\downarrow} (p^\mu) + m^2
\rho^A_{\uparrow} (p^\mu) = 0 \quad,\\
\gamma_{\mu\nu} p^\mu p^\nu \lambda^A_{\rightarrow} (p^\mu) - m^2
\lambda^A_{\rightarrow} (p^\mu) &=& 0 \quad,\quad
\gamma_{\mu\nu} p^\mu p^\nu \rho^A_{\rightarrow} (p^\mu) - m^2
\rho^A_{\rightarrow} (p^\mu) = 0 \quad .
\end{eqnarray}

There exist the identities analogous to (\ref{i1},\ref{i2}).
For instance,   under the choice of the phase factors
as $\delta_1^R = \delta_3^R = 0$,
$\delta_1^L = \delta_3^L=0$ and $\delta_2^L = \delta_2^R -\pi =0$ we
have\footnote{{\it Cf.} with the formulas (21a-c) in [22a]
and (48,49) in ref.~\cite{DVOax}.
Thus, the form of these relations depends on the choice of the
spinorial basis and it is governed by the covariance of the theory
under discrete symmetries.}
\begin{eqnarray}
\rho^S_{\uparrow} (p^\mu) &=& +\lambda^S_\downarrow (p^\mu)\quad,\quad
\rho^S_\downarrow (p^\mu)= +\lambda^S_\uparrow (p^\mu)
\quad,\quad \rho^S_\rightarrow (p^\mu)= -
\lambda^S_\rightarrow (p^\mu)\quad,\\
\rho^A_{\uparrow} (p^\mu)&=& - \lambda^A_\downarrow (p^\mu)\quad,\quad
\rho^A_\downarrow (p^\mu) = -\lambda^S_\uparrow (p^\mu)\quad,\quad
\rho^A_\rightarrow (p^\mu)= +\lambda^A_\rightarrow (p^\mu)\quad .
\end{eqnarray}
Therefore,
\begin{eqnarray}
\gamma_{\mu\nu} p^\mu p^\nu \lambda^S_{\uparrow\downarrow\rightarrow}
(p^\mu) - m^2 \rho^S_{\uparrow\downarrow\rightarrow} (p^\mu)&=&0\quad,\quad
\gamma_{\mu\nu} p^\mu p^\nu \lambda^A_{\uparrow\downarrow\rightarrow}
(p^\mu) - m^2 \rho^A_{\uparrow\downarrow\rightarrow} (p^\mu)=0\quad,\\
\gamma_{\mu\nu} p^\mu p^\nu \rho^S_{\uparrow\downarrow\rightarrow} (p^\mu)
- m^2 \lambda^S_{\uparrow\downarrow\rightarrow} (p^\mu)&=&0\quad,\quad
\gamma_{\mu\nu} p^\mu p^\nu \rho^A_{\uparrow\downarrow\rightarrow} (p^\mu)
- m^2 \lambda^A_{\uparrow\downarrow\rightarrow} (p^\mu)=0\quad.
\end{eqnarray}
Applying relations between type-II and type-I spinors that look like
similar to (\ref{con1}-\ref{conl})
except for $\rho^S \leftrightarrow \rho^A$ we obtain
\begin{eqnarray}\label{eqb}
\left (\gamma_{\mu\nu} p^\mu p^\nu - m^2 \right )\psi^D (p^\mu) +
\left (\gamma_{\mu\nu} p^\mu p^\nu - m^2 \right )\gamma_5(\psi^D (p^\mu))^c
&=&0\quad,\quad\\
\left (\gamma_{\mu\nu} p^\mu p^\nu + m^2 \right )\gamma_5\psi^D (p^\mu) -
\left (\gamma_{\mu\nu} p^\mu p^\nu + m^2 \right )(\psi^D (p^\mu))^c
&=&0\quad,\quad\\
\left (\gamma_{\mu\nu} p^\mu p^\nu - m^2 \right )\psi^D (p^\mu) -
\left (\gamma_{\mu\nu} p^\mu p^\nu - m^2 \right )\gamma_5(\psi^D (p^\mu))^c
&=&0\quad,\quad\\ \label{eqbl}
\left (\gamma_{\mu\nu} p^\mu p^\nu + m^2 \right )\gamma_5\psi^D (p^\mu) +
\left (\gamma_{\mu\nu} p^\mu p^\nu + m^2 \right )(\psi^D (p^\mu))^c
&=&0\quad.\quad
\end{eqnarray}
This  tells us that $\psi^D$ (or $\gamma_5 (\psi^D)^c$)
should be considered as the positive-energy solutions
of the modified Weinberg equation~\cite{DVA0}
and $(\psi^D)^c$ (or $\gamma_5 \psi^D$) as the negative-energy ones.
The analogs of  the equations (\ref{cr1},\ref{cr2}) can be written:
\begin{eqnarray}
\gamma^{\mu\nu} \partial_\mu \partial_\nu \lambda_\eta (x)
+ \wp_{S,A} m^2 \lambda_{-\eta} (x) &=& 0\quad,\\
\gamma^{\mu\nu} \partial_\mu \partial_\nu \rho_\eta (x)
+ \wp_{S,A} m^2 \rho_{-\eta} (x) &=& 0\quad,
\end{eqnarray}
where $\wp_{S,A} =\pm 1$, the sign is ``$+$"
for positive-energy solution $\lambda^S (p^\mu)$ (or $\rho^S (p^\mu)$)
and the sign is ``$-$" for negative-energy solutions $\lambda^A (p^\mu)$
(or $\rho^A (p^\mu)$).
This refers to the $\eta =\uparrow$ or $\eta=\downarrow$. As
for  $\eta=\rightarrow$ it is easy to see that
the equations (85,88) have the opposite signs at mass terms.

The presence of
$\wp_{\uparrow\downarrow}$ in a $j=1/2$ case
or $\wp_{S,A}$ in a $j=1$ case hints that we obtained the examples of
the FNBWW-type quantum field theory. The analysis
of the field operator
(\ref{fo}) in the Fock space reveals that fermion
and its antifermion can possess same intrinsic
parities~[61,22d].
Bosons described by the Eqs. (\ref{eqb}-\ref{eqbl})
are found following to ref.~\cite{DVA0}
to be able to carry opposite intrinsic parities, depending on
the choice of the field operator.

\section{Concluding Remarks}

In this paper I have presented an overview of the theory of truly neutral
particles. The question of  applicability of the new constructs
in the $(j,0)\oplus (0,j)$ representation space to neutrino
physics has been discussed. The connection of the new models with the
theories envisaged long ago by Foldy and Nigam~\cite{Nigam},
and Bargmann, Wightman and Wigner~\cite{Wigner2} has been
found. The particle properties with respect to the operation
of parity, being discussed in the present
paper (and in refs.~\cite{DVA0,DVA}), are unusual. In fact, it was
shown that these properties depend on the choice of the field operator.
Moreover, it was found that the physical content depends on the choice
of the spinorial basis.
Research in the framework of other constructs
representation space deserves further elaboration.

Unfortunately, the present experimental data don't yet permit us
to make reliable conclusions on the sufficiency of the Standard
Model (and to what limits). However, the wide interest in neutrino
physics in the theoretician community
and  the forthcoming experimental facilities,
SUPER-KAMIOKANDE, SNO (Sudbury),
BOREXINO, ICARUS (CERN-Gran Sasso), HELLAZ, HERON (see,
{\it e.~g.}, the proceedings of the recent neutrino
conference~\cite{experim}),
leave us with a hope that the puzzles of misterious
neutral particles can be resolved in a short time.

{\it Acknowledgements.} I greatly appreciate
many helpful advises of Prof. D. V. Ahluwalia,
Prof. A. F. Pashkov and  Prof. Yu. F. Smirnov.
Discussions with Profs. I. G. Kaplan and M. Torres on
neutrino experiments were very useful.
The questions of Prof. M. Moreno and Prof. A. Turbiner
helped me to realise the necessity of the study of neutral
particles.

I would also like to thank an unknown referee of ``Phys. Lett. B",
who found that the conclusion of the papers~\cite{DVO} is
``a known result".  His report was in a certain part encouraging
and it gave me an additional impulse to speed up
the work in the needed direction, what resulted
in the writing of the present paper.

I am grateful to Zacatecas University for a professorship.

\nopagebreak

\end{document}